\documentstyle[12pt,a4wide]{article}
\makeatletter

\@ifundefined{chapter}{\def\thebibliography#1{\section*{References\@mkboth
  {REFERENCES}{REFERENCES}}\list
  {\relax}{\setlength{\labelsep}{0em}
	\setlength{\itemindent}{-\bibhang}
	\setlength{\itemsep}{0pt}
	\setlength{\parsep}{0pt}
	\setlength{\leftmargin}{\bibhang}}
    \def\newblock{\hskip .11em plus .33em minus .07em}
    \sloppy\clubpenalty4000\widowpenalty4000
    \sfcode`\.=1000\relax}}%
{\def\thebibliography#1{\chapter*{Bibliography\@mkboth
  {BIBLIOGRAPHY}{BIBLIOGRAPHY}}\list
  {\relax}{\setlength{\labelsep}{0em}
	\setlength{\itemindent}{-\bibhang}
	\setlength{\itemsep}{0pt}
	\setlength{\parsep}{0pt}
	\setlength{\leftmargin}{\bibhang}}
    \def\newblock{\hskip .11em plus .33em minus .07em}
    \sloppy\clubpenalty4000\widowpenalty4000
    \sfcode`\.=1000\relax}}

\newlength{\bibhang}
\let\@internalcite\cite
\def\cite{\@ifstar{\citeyear}{\citefull}}
\def\citefull{\def\astroncite##1##2{##1 ##2}\@internalcite}
\def\citeyear{\def\astroncite##1##2{##2}\@internalcite}

\def\@citex[#1]#2{\if@filesw\immediate\write\@auxout{\string\citation{#2}}\fi
  \def\@citea{}\@cite{\@for\@citeb:=#2\do
    {\@citea\def\@citea{; }\@ifundefined
       {b@\@citeb}{{\bf ?}\@warning
       {Citation `\@citeb' on page \thepage \space undefined}}%
{\csname b@\@citeb\endcsname}}}{#1}}

\def\@cite#1#2{(#1\if@tempswa , #2\fi)}
\def\@biblabel#1{}
\makeatother

\newcommand{\etal}{et~al.}
\newcommand{\ionhy}{H{\sc ii}}
\newcommand{\water}{$\mbox{H}_{2}\mbox{O}$}
\newcommand{\methanol}{$\mbox{CH}_{3}\mbox{OH}$}
\newcommand{\formaldehyde}{$\mbox{H}_{2}\mbox{CO}$}
\newcommand{\transa}{$5_{1}\mbox{-}6_{0}\mbox{~A}^{+}$}
\newcommand{\transe}{$2_{0}\mbox{-}3_{-1}\mbox{~E}$}

\newcommand{\kms}{$\mbox{km~s}^{-1}$}

\title{
  A Search for Extragalactic Methanol Masers}
\author{ \bigskip \\
  S.P. Ellingsen,$^{1}$ R.P. Norris,$^{2}$ J.B. Whiteoak,$^{2}$
  R.A. Vaile,$^{3}$ \\ P.M. McCulloch$^{1}$ and M.Price$^{2}$}

\begin{document}

\maketitle

1 - Department of Physics, University of Tasmania, GPO 252C, Hobart, TAS 7001,
    Australia. \\ [4mm]

2 - Australia Telescope National Facility, CSIRO Radiophysics Laboratory,
    PO~Box~76, Epping, NSW 2121, Australia. \\ [4mm]

3 - Department of Physics, University of Western Sydney, Macarthur,
    PO~Box~ 555, Campbelltown 2560, Australia \\ [4mm]

Key Words : Masers, Megamasers, Methanol

\section*{Abstract}

A sensitive search for 6.7--GHz methanol maser emission has been made towards
10
galaxies that have yielded detectable microwave molecular--line transitions.
These include several which show OH megamaser or superluminous \water\/ maser
emission.  Within the Galaxy, \methanol\/ and OH masers often occur in the same
star formation regions and, in most cases, the \methanol\/ masers have a
greater
peak flux density than their OH counterparts.  Thus we might expect
\methanol\/ masers to be associated with extragalactic OH maser sources.
We failed to detect any emission or absorption above our 60--mJy
detection limit.  We conclude that if the physical conditions exist to
produce \methanol\/ megamaser emission, they are incompatible with the
conditions which produce OH megamaser emission.

\section*{Introduction}

OH maser emission in galaxies other than our own was first detected
20 years ago \cite{Wh1973}.  Since then, extragalactic maser emission
has been detected from three other molecules, \water, \formaldehyde\/ and CH
\cite{Ch1977,Ba1986,Ba1990,Wh1980}. In most cases the
extragalactic masers seem to be far stronger versions of Galactic masers.
However, the OH molecule has also been found to exhibit a different class of
maser emission.  Megamaser galaxies, first discovered by Baan, Wood \& Haschick
\cite*{Ba1982}, are
galaxies in which a substantial fraction of the molecular gas surrounding the
nucleus is stimulated to emit maser radiation, so that the galaxy as a whole
appears as a maser some million times more luminous than a normal Galactic
maser.  Where they occur, \formaldehyde\/ emission appears to be associated
with OH megamasers, and CH with superluminous \water\/ masers.

In recent years, strong maser emission in our Galaxy has been discovered in
the 12.2--GHz \transe\/ and 6.7--GHz \transa\/ transitions of \methanol\/
\cite{Ba1987,Me1991}.  So far, maser emission has only been
detected towards star-formation regions, although absorption has been detected
towards both these regions and cold clouds \cite{Wa1988,Pe1991}.  \methanol\/
masers appear to be closely associated with OH and \water\/ masers
\cite{No1993,Me1993} and, in at least one case, the \methanol\/ and OH masers
are coincident within 2 arcsec, or 4000 AU. Since the same conditions appear
to produce both OH and \methanol\/ masers, we might expect OH megamaser
emission to be accompanied by detectable \methanol\/ maser or megamaser
emission.

A preliminary search for the 12.2--GHz \methanol\/ transition was made by
Norris
\etal\/ \cite*{No1987} towards a few known OH megamaser galaxies, but no
systematic sensitive search has been published so far. Here we present the
results of an exploratory search, made at the stronger 6.7--GHz transition.

\section*{Observations}

The observations were made between 1992 February 25 and March 9 using the
dual--channel cooled HEMT 6.7/12.2--GHz receiver at the Parkes 64--m telescope
which, at 6.7~GHz, has a beamwidth of 3.3~arcmin.  The equivalent system
temperature
for the observations was $\sim60$~K. An autocorrelator provided two
512--channel spectra in orthogonal linear polarizations, each spread over
64~MHz.  Thus the observations covered a velocity extent of $\sim2800$~\kms\/
and had a velocity resolution (after Hanning smoothing) of
7.8~\kms\/.  The spectra were obtained by taking two 10--min spectra
on--source and two reference spectra, one offset by +15~min and the other
by -15~min of right ascension.  These were then used to produce two
quotient spectra each with different references, yielding a total on--source
time of 20~min.  The resulting spectra for the two polarizations were
then averaged and Hanning smoothed.  The resulting rms noise level of a
10--min observation was typically 0.04~Jy. To achieve the desired sensitivity,
multiple observations were made of each source. The total integration time for
most sources was 20--40 min, but sometimes exceeded 1 hr; the
resulting $3\sigma$ detection level was typically no greater than 0.06~Jy.

Flux density calibration was carried out using observations of the sources
PKS~0407-658, Hydra~A and PKS~1934-638, which were assumed to have flux
densities of 2.19, 4.09 and 9.84~Jy respectively.

Ten galaxies were surveyed for the \transa\/ \methanol\/ transition ; six
are known OH maser or megamaser sources \cite{Wh1973,No1989,Ka1990} and two are
known superluminous \water\/ masers \cite{Wh1986}.  Thus the sample is strongly
biased towards galaxies which show ultraluminous maser emission in other
transitions.

\section*{Results}

Our results are shown in Table 1. None of the ten galaxies observed contains a
detectable 6.7-GHz \methanol\/ maser.

The detection threshold of 60~mJy is significantly lower than that necessary to
detect the OH and \water\/ masers which exist in these sources. Given that the
Galactic \methanol\/ masers are typically much stronger than Galactic OH
masers, this detection limit places a severe  constraint on any \methanol\/
maser emission, and demonstrates that the OH megamaser emission and
superluminous \water\/ maser emission are not accompanied by corresponding
\methanol\/ megamaser emission.

\section*{Discussion}

In our Galaxy, 6.7-GHz \methanol\/ masers are found solely in star formation
regions, and are closely associated with OH and \water\/ masers. Existing
surveys (J. L. Caswell et al. in preparation) indicate that nearly all known OH
masers are accompanied
by 6.7--GHz \methanol\/ activity, and vice-versa.
6.7--GHz \methanol\/ masers
have been detected towards two \ionhy\/ regions in the Large Magellanic Cloud
\cite{Si1992}, S. P. Ellingsen et al. in preparation).
The intrinsic peak flux density of these sources
is of a similar strength to most Galactic masers.  Since the same conditions
appear to produce both OH and \methanol\/ masers within the galaxy, we might
expect the OH maser
and megamaser emission in other galaxies to be accompanied by detectable
\methanol\/ maser emission, possibly even \methanol\/ megamaser emission.
Furthermore, in Galactic sources, the 6.7-GHz \methanol\/ maser emission is
typically much stronger than the corresponding OH emission, and so we might
even expect extragalactic \methanol\/ maser emission to be much stronger than
that of the OH emission.
The ratio of the peak flux densities of 6.7--GHz
\methanol\/ and OH masers spans several orders of magnitude, but is typically
of the order of ten.  With one exception, the sensitivity of this search would
have been sufficient to detect any 6.7--GHz \methanol\/ with peak flux density
comparable to the OH or \water\/ sources observed in these galaxies
(see table 1).  If we assume that in our sample there are no 6.7--GHz
\methanol\/
sources with peak flux greater than 3 times the quoted RMS noise level, then we
have four sources with \methanol\/ : OH flux ratios less than $0.3$.
Among Galactic masers, approximately 23\% have \methanol : OH flux ratios less
then $0.3$, thus if we assume the same \methanol\/ : OH flux ratio distribution
for extragalactic sources then the probability that any four will all have
flux ratios less than $0.3$, is 0.28\%.  Hence it appears extremely unlikely
that the extragalactic \methanol\/ : OH flux ratio distribution is the same
as that observed for Galactic masers.

The differences between Galactic and extragalactic masers sources might be
attributed to one of the following causes.

\newcounter{listcount}

\begin{list}
{(\roman{listcount})}{\usecounter{listcount}\setlength{\rightmargin}{\leftmargin}}

\item \methanol\/ megamasers do not exist, because the physical conditions
      required to produce them do not exist.

\item Extragalactic \methanol\/ masers or megamasers do exist, but require
      different physical conditions from those which produce ultraluminous
      OH and \water\/ maser emission.

\item The pumping mechanism or efficiency of \methanol\/ masers is such that
      peak flux density of extragalactic \methanol\/ masers is below the
detection
      limit of these observations.

\end{list}

Megamaser emission requires a number of basic ingredients, such as a sufficient
column density of molecules along the line of sight, a means of pumping
the masers, and perhaps a background continuum source
to provide the input to the maser.  Galactic masers appear to  require precise
physical conditions such as a particular optical
depth to the pump radiation.  However, megamasers are relatively insensitive
to the precise conditions, because the maser activity in these sources is
distributed throughout a large region, and a wide range of
physical conditions are available if the basic ingredients are present.

Thus our first hypothesis, that \methanol\/ megamasers do not exist because the
physical conditions required to produce them do not exist, implies that
some physical condition is required for methanol maser emission, but that this
condition is found only in very special circumstances, and will not be
widespread through the disc of a galaxy. An example might be if
Galactic \methanol\/ masers occur only in concentrations of high density within
protoplanetary discs, as suggested by Norris \etal\/ \cite*{No1993}. It is
possible that the
mechanisms which produce increased \methanol\/ density in Galactic star
formation regions \cite{He1991} cannot operate on a sufficiently large scale,
or the radiation field in these regions causes depletion by disassociation
of the \methanol\/ molecules.

Our second hypothesis, that \methanol\/ megamasers do exist, but require
different physical conditions from those of OH megamasers, would be
appropriate if, for example, the \methanol\/ masers were radiatively pumped
but the OH megamasers collisionally pumped. However, detailed differences,
such as optical--depth effects, would not be sufficient to prevent megamaser
emission.

The final hypothesis, that the extragalactic methanol masers are below the
detection limit of our observations, implies that either the peak flux density
of \methanol\/ masers cannot greatly exceed that of the strongest Galactic
\methanol\/ masers, or the conditions which produce ultraluminous Galactic
type OH and \water\/ masers are not suitable for producing ultraluminous
\methanol\/ masers.  We cannot attribute the non--detection of extragalactic
masers to a deficiency of \methanol, as it has been detected towards several
galaxies at millimetre wavelengths \cite{He1987}.  One
of the galaxies which we also observed (NGC~253), was found to have methanol
abundances similar to those found in our Galaxy. NGC~253 is also the closest
of the observed galaxies, but to detect any masers in our observations,
an intrinsic peak flux density at least an order of magnitude
greater than the strongest of the Galactic \methanol\/ masers would have been
required.

All of these cases place a severe constraint on models of \methanol\/
maser emission. To determine whether extragalactic \methanol\/ masers are
common requires a more sensitive and more comprehensive survey.

\section*{Conclusion}

We conclude that the absence of \methanol\/ maser or megamasers implies that
either the physical conditions required to produce ultraluminous \methanol\/
maser emission are incompatible with those required to produce OH or \water\/
emission, or that the ingredients necessary to produce masing in
\methanol\/ are not present on a large enough scale to produce megamaser
emission.

\bibliographystyle{mnras}

\newpage

\parskip 0mm
\parindent 0mm

{\bf Table 1.} The selected sample of galaxies. References : a, Whiteoak \&
Gardener \cite*{Wh1986}; b, Staveley--Smith \etal\/ \cite*{St1992};
c, Baan, Wood \& Haschick \cite*{Ba1982}; d, Bottinelli \etal\/ \cite*{Bo1987};
e, Whiteoak \& Gardner \cite*{Wh1973}; f, Norris \etal\/ \cite*{No1989};
g, L\'{e}pine \& dos Santos \cite*{Le1977}; h, Dos Santos \& L\'{e}pine
\cite*{Do1979}; i, Claussen, Heiligman \& Lo \cite*{Cl1984}


\begin{tabbing}
xxxxxxxxxx \= xxxxx \= xxxxxxxx \= xxxxxxxxxx
\= xxxxxxxxxxxx \= xxxxxxxxxx \= xxxxxxxxx \= xxxxxxxx \= \kill
{}~~Source            \>            \> Position (B1950) \>       \> ~~~Velocity
\> RMS Flux  \> OH Peak \> \water Peak \> Ref. \\
                   \> Right Ascension \> \> Declination \> range (\kms)
\> ~~Density \> ~~Flux   \> ~~Flux \\
                      \> ~~~h~~m~~s     \> \> ~~~$^{\circ}$~~~$'$~~~$''$ \>
\> ~~~(Jy)   \> ~~(Jy)  \> ~~(Jy) \\
\end{tabbing}


\parskip 0mm

\begin{tabbing}
YYYYYYYY \= hh mm ssxxxxx \= sdd mm ssx \= yyyyxxyyyxxxx
\= y.yyyxxxxx \= y.yyyxxxx \= yy--yyxxxx \= x \kill
NGC 253    \> 00 45 06 \> -25 34 00 \> -900$\rightarrow$1400
\> 0.01 \> 0.120 \> ~~5               \> b,e,g \\
NGC 1068   \> 02 40 07 \> -00 13 30 \> ~300$\rightarrow$2500
\> 0.03 \> ~~~-- \> ~0.7              \> i \\
NGC 1487   \> 04 04 05 \> -42 30 42 \> -300$\rightarrow$1900
\> 0.008 \> ~~~--\> ~~~--             \> \\
NGC 1566   \> 04 18 53 \> -55 03 24 \> ~400$\rightarrow$2600
\> 0.01 \> $<0.040$ \> ~~~--             \> f \\
10039-3338 \> 10 03 55 \> -33 38 43 \> 9000$\rightarrow$11200
\> 0.02 \> 0.315 \> ~~~--             \> b \\
11506-3851 \> 11 50 40 \> -38 51 10 \> 2000$\rightarrow$4200
\> 0.01 \> 0.105 \> ~~~--             \> b \\
NGC 4418   \> 12 24 23 \> -00 36 14 \> 1100$\rightarrow$3400
\> 0.02 \> 0.004 \> ~~~--             \> d \\
NGC 4945   \> 13 02 32 \> -49 12 02 \> -600$\rightarrow$1700
\> 0.02 \> -0.800 \> 9$\rightarrow$16  \> a,b,e,h \\
Circinus   \> 14 09 18 \> -65 06 19 \> -600$\rightarrow$1700
\> 0.02 \> ~~~-- \> 3$\rightarrow$12  \> a \\
Arp 220    \> 15 32 47 \> ~23 40 10 \> 4300$\rightarrow$6500
\> 0.01 \> 0.280 \> ~~~--             \> c \\
\end{tabbing}

\end{document}